\documentclass[%
 reprint, onecolumn,
 amsmath, amssymb,
 aps,
]{revtex4-2}

\usepackage{graphicx}
\usepackage{dcolumn}
\usepackage{bm}
\usepackage{xcolor}
\usepackage{amsmath}
\usepackage{mathrsfs}

\begin{document}
\preprint{AIP/123-QED}
\title{Electric field effects on the collision efficiency of uncharged water droplets in a linear flow}

\author{Pijush Patra}
\affiliation{Nordita, KTH Royal Institute of Technology and Stockholm University, Stockholm 10691, Sweden}
\author{Anubhab Roy}
\affiliation{Department of Applied Mechanics and Biomedical Engineering, Indian Institute of Technology Madras, Chennai, Tamil nadu 600036, India}
\author{J. S. Wettlaufer}
\homepage{Author to whom correspondence should be addressed: jw@fysik.su.se, john.wettlaufer@yale.edu}
\affiliation{Nordita, KTH Royal Institute of Technology and Stockholm University, Stockholm 10691, Sweden and Departments of Applied Mathematics, Earth \& Planetary Sciences, and Physics, Yale University, New Haven, Connecticut 06520-8109, USA}

\begin{abstract}
We study the dynamics of collisions between a pair of uncharged conducting droplets under the influence of a uniaxial compressional flow and an external electric field. The near-field asymptotic expression for the electric-field-induced attractive force demonstrate that surface-to-surface contact in finite time is facilitated by overcoming lubrication resistance. We demonstrate the significant role of the external electric field on the relative trajectories of two droplets in a compressional flow and provide estimates of the correlation between collision efficiency and the forces induced by the electric field. For droplet collisions in clouds, continuum lubrication approximations become inadequate to capture collision dynamics, and thus we incorporate non-continuum lubrication interactions into our analysis to address this complexity. Our findings reveal the dependence of collision efficiency on the strength of the electric field, geometry of the two interacting droplets, non-continuum effects, and van der Waals forces.
\end{abstract}

\maketitle

\section{Introduction}\label{Introduction}

Understanding the mechanisms of droplet growth that lead to the rapid onset of warm rain poses a significant challenge in cloud microphysics \citep{grabowski2013growth}. According to classical condensation theory, a cloud droplet's growth rate varies inversely with its radius, making it unlikely that condensation alone can produce raindrops within a realistic time scale \citep{jonas1996turbulence}. Furthermore, condensation tends to result in a nearly monodisperse, narrow droplet size distribution (DSD), while in situ measurements reveal a broader DSD \citep{prabha2011microphysics,khain2013mechanism}. Therefore, collisions and subsequent coalescence between droplets are responsible for forming larger droplets that lead to warm rain initiation \citep{pruppacher1997microphysics}. When droplet radii exceed approximately 30 \textmu m, 
gravity-induced collisions accelerate the growth process, but for smaller droplets gravitational collisions are not efficient. Turbulent airflow is vital in promoting droplet collisions, especially for equal-sized droplets. Many researchers have primarily focused on the role of gravitational sedimentation and turbulence in droplet collisions (see \cite{shaw2003particle,devenish2012droplet,grabowski2013growth} and references therein). However, cloud droplets are not just passive particles; they often carry electric charges and are affected by electric fields within clouds. Despite this, relatively few studies have explored the impact of electrostatic forces on the collision-coalescence process \citep{tinsley2000effects,khain2004rain,lu2015charged,boutsikakis2022effect,ruan2024effects,patra2023collision,dubey2024critical}. These electrostatic forces may play a central role in the early stages of cloud development, where collisions between small droplets create size differences in the droplet spectrum essential for continuing the growth process \citep{poydenot2024pathways}. More importantly, incorporating an accurate parameterization of electrostatic forces into large-scale models, such as numerical weather prediction simulations, could significantly improve forecasting accuracy. Here we study how electrostatic forces arising from an external electric field influence droplet collisions.

A downward pointing fair-weather electric field arises from the potential difference between Earth's surface and the upper atmosphere, causing clouds to become electrified \citep{pruppacher1997microphysics,wang2013physics}. In thunderclouds, the electric field rapidly intensifies due to charge separation driven by cloud charging mechanisms, such as convective, inductive, and non-inductive processes, a thorough discussion of which is given in Sec. 18.5 of \citet{pruppacher1997microphysics} and Sec. 14.4 of \citet{wang2013physics}. Laboratory experiments have highlighted the significance of collisional charging during ice-ice collisions in thundercloud electrification \citep{reynolds1957thunderstorm,latham1965electrification,takahashi1978riming, gaskell1980charge,jayaratne1983laboratory,mason2000charge,saunders2006laboratory,turner2022effects}, which has inspired the development of theoretical models aimed at explaining this phenomenon \citep[e.g.,][]{baker1987, baker1989charge,Williams1991,jayaratne1993,baker1994mechanism,dash2001theory,dash2003surface,jungwirth2005possible}.

The electric field strength in clouds during fair-weather conditions ranges from about $10^2 - 10^3$ V/m, whereas in moderately to highly electrified clouds it typically ranges from \(10^4\) to \(10^5\) V/m (See e.g., \citet{gunn1948electric}; Chap. 18, pp. 804-811 of \citet{pruppacher1997microphysics}; Chap. 3, pp. 86-87 of \citet{rakov2003lightning}; \citet{trinh2020determining}). \citet{winn1974measurements} documented instances where the field strength reached \(4 \times 10^5\) V/m. Moreover, these studies indicate that the electric field can be oriented from normal to parallel to the direction of gravity.
Wind tunnel experiments have demonstrated that strong electric fields can significantly alter the impact velocities and shapes of droplets, affecting their collision characteristics \citep{bhalwankar2007wind,bhalwankar2009wind,bhalwankar2023binary,pawar2024effects}. More importantly, in their field observations, \citet{mudiar2018quantification,mudiar2021electric} reported that these electric fields in thunderclouds promote raindrop growth and rainfall rates. Therefore, these observations motivate theoretical investigations into the role of electric fields in droplet collision dynamics.

An applied electric field induces the build up of opposite charges on the adjacent sides of two neutral conductive droplets. The interactions between these induced charges amplify the local electric field between the droplets as they approach one another, resulting in an increase in the attractive electrostatic force \citep{lekner2011near}. This force increases without bound as the droplets converge, facilitating surface-to-surface contact by overcoming lubrication resistance \citep{lekner2013forces}. However, at larger separations the electric force diminishes, making contact unlikely unless background flows, gravity, and/or thermal fluctuations drive relative droplet motions. 

The electrostatic force between two spherical conductors in an external electric field has been studied extensively. \citet{davis1964two} calculated this force by integrating electrical stresses over the surfaces of each sphere, which depends on the relative sizes of the two spheres, the amount of surface charge, and the electric field strength. The formulation involves an infinite series, which converges extremely slowly for small separation distances. \citet{lekner2013forces} overcame this challenge by expressing these forces in terms of polarizabilities. To obtain the electric-field-induced force on the close approach of two arbitrarily-sized spherical conductors, we utilize the work of \citet{lekner2011polarizability}, who derived the exact analytical expressions for the longitudinal and transverse polarizabilities for small separation distances. Recently, \citet{thiruvenkadam2023pair} analyzed the relative trajectories of two uncharged conducting spheres of arbitrary sizes in an external electric field alone. They demonstrated that, due to the divergent nature of the electric field-induced forces in the lubrication region, these spheres could come into contact in a finite time. Here we include the exact lubrication form of electric-field-induced forces.

In their pioneering work, \citet{saffman1956collision} calculated the collision rate of droplets in a cloud setting by approximating the turbulent flow experienced by the sub-Kolmogorov droplets as a pseudo-steady uniaxial compressional flow with Gaussian statistics for strain rates. They found the collision rate to be $(8 \pi /15)^{1/2} n_1 n_2 \varGamma_{\eta} (a_1+a_2)^3$, where $a_1$ and $a_2$ are the droplet radii with number densities $n_1$ and $n_2$ respectively, $\varGamma_{\eta}=(\epsilon/\nu_f)^{1/2}$ is the Kolmogorov shear rate, with $\epsilon$ the turbulent energy dissipation rate and $\nu_f$ the kinematic viscosity of the surrounding fluid. In clouds, droplet sizes are significantly smaller than the Kolmogorov length scale ($\sim 1$ mm) of turbulence. Consequently, the Reynolds number based on the droplet length scale is much less than unity. In this scenario, researchers have studied droplet collisions in turbulence by approximating the background turbulent flow in the vicinity of a droplet pair as a stochastically varying linear flow \citep{brunk1998turbulent,chun2005clustering,dhanasekaran2021turbulent}. The most common realization of this stochastic flow is a compressional flow \citep{ashurst1987alignment}. \citet{zeichner1977use} determined the collision rate of two non-interacting spherical droplets subject to a steady uniaxial compressional flow to be $(8 \pi /3\sqrt{3})n_1 n_2 \dot{\gamma} (a_1+a_2)^3$, where $\dot{\gamma}$ is the compression rate. Here, we consider the background flow as a frozen uniaxial compressional flow. We can express the compression rate in terms of Kolmogorov shear rate by equating the collision rates evaluated by \citet{saffman1956collision} and \citet{zeichner1977use}, yielding $\dot{\gamma}=(9/40\pi)^{1/2}(\epsilon/\nu_f)^{1/2}$. The turbulent energy dissipation rate depends on the cloud type. For example, $\epsilon \sim 10^{-1}$ m$^2$s$^{-3}$ in cumulonimbus clouds (highly turbulent clouds), $\epsilon \sim 10^{-2}$ m$^2$s$^{-3}$ in cumuli and $\epsilon \sim 10^{-3}$ m$^2$s$^{-3}$ in stratocumuli (low turbulence clouds). Hence, for $\nu_f \approx 10^{-5}$ m$^2$s$^{-1}$, the compression rate range is $\dot{\gamma} \sim 1-30$ s$^{-1}$.

The inertia of larger droplets can significantly impact collision dynamics, but this effect is small enough to be neglected for relatively small droplets. Thus, we assume that fluid inertia is negligible, allowing us to use the Stokes equations to describe the flow field. Additionally, we assume that the droplets are sufficiently large that thermal diffusion is negligible. We justify these assumptions by calculating the relevant dimensionless parameters for a water droplet (density $\rho_p \approx 10^3$ kg m$^{-3}$) of radius $a_1=10$ \textmu m in air (dynamic viscosity $\mu_f \approx 1.8 \times 10^{-5}$ Pa-s and density $\rho_f \approx 1$ kg m$^{-3}$) at temperature $T=275$ K. For a typical compression rate $\dot{\gamma}=25$ s$^{-1}$, the droplet Reynolds number is $Re_p=\rho_f\dot{\gamma} a_1^2/\mu_f \approx 1.38 \times 10^{-4}$, the Stokes number is $St=2 a_1^2\rho_p\dot{\gamma}/9\mu_f \approx 0.03$, and the Peclet number is $Pe=3\pi \mu_f \dot{\gamma}a_1^3/k_BT \approx 1.12 \times 10^3$, where $k_B=1.318 \times 10^{-23}$JK$^{-1}$ is Boltzmann's constant. These values of $Re_p$, $St$ and $Pe$ clearly support our assumptions.

The characteristic hydrodynamic and electric stresses for a spherical water droplet subjected to a uniaxial compressional flow and an electric field are given by $\mu_f \dot{\gamma}$ and $\epsilon_0 E_0^2$, respectively, where $\epsilon_0 = 8.85 \times 10^{-12}$ Fm$^{-1}$ is the free space permittivity (ostensibly that of air) and $E_0$ is the electric field strength. The flow capillary number measures the relative strength of hydrodynamic and capillary stresses, and is defined as $Ca = \mu_f \dot{\gamma} a_1 /\sigma$, where $\sigma \approx 72 \times 10^{-3}$ Nm$^{-1}$ is the air-water surface tension. Similarly, the electric capillary number is the ratio of the electric stress to interfacial tension stress, and is defined as $Ca_E = \epsilon_0 E_0^2 a_1/\sigma$. The deformation of a droplet from a spherical shape depends on the magnitudes of these two parameters. For $a_1 = 10$ \textmu m, $\dot{\gamma} = 25$ s$^{-1}$, and $E_0 = 10^5$ Vm$^{-1}$, we find that $Ca \approx 6.25 \times 10^{-8}$ and $Ca_E \approx 1.23 \times 10^{-5}$, which are both sufficiently small that we neglect droplet deformation. Furthermore, interfacial fluctuations becomes insignificant due to the high droplet-to-medium viscosity ratio ($\sim 10^2$) for water droplets in air. Therefore, it is a very good approximation to treat small droplets as rigid spheres. 

When two droplets approach each other, they create disturbance flow fields in the host fluid medium that enhance the hydrodynamic resistance on each droplet. 
This hydrodynamic interaction can significantly alter the outcome of collisions. We quantify this effect by calculating the collision efficiency, which is the ratio of the collision rate with and without hydrodynamic interactions. There is extensive treatment in the literature of two-body hydrodynamic interactions in Stokes flow (See e.g., the book by \citet{kim2013microhydrodynamics}). The nondimensional surface-to-surface separation distance is $\xi = [r-(a_1+a_2)]/a^* = (r/a^*)-2$, where $r$ and $a^*=(a_1+a_2)/2$ are the center-to-center distance between the droplets and the mean radius of the two droplets
respectively.  In the lubrication region, the hydrodynamic resistance due to the normal motion of two surfaces is O$(1/\xi)$, indicating that droplets will not come into contact in finite time. 
For very small droplets, the continuum lubrication approximation would not be valid when the separation distance between the surfaces is less than the mean free path of air molecules $\lambda_0$. Therefore, in such a situation we consider non-continuum lubrication resistance, which is O$(1/[\ln(\ln(Kn/\xi))/Kn])$ \citep{sundar96non}. 
Here, $Kn=\lambda_0/a^*$ is the Knudsen number, which measures the significance of non-continuum interactions.
This non-continuum lubrication resistance has a weaker divergence than its continuum counterpart, which allows for collisions in finite time.  Previously, researchers have quantified the effects of non-continuum interactions on the collision rate of droplets subject to thermal fluctuations \citep{patra2022brownian}, laminar flows \citep{dhanasekaran2021collision,patra_koch_roy_2022}, and a turbulent flow \citep{dhanasekaran2021turbulent}.

This paper is structured as follows. We start by defining the problem and describing the method for computing the collision rate and efficiency in Sec. \ref{Problem_formulation}. In Sec. \ref{Results_and_discussion}, we show the effects of a background linear flow, an external electric field, and van der Waals forces on collision efficiency. We summarize our results and conclude in Sec. \ref{Conclusions}.

\section{Problem Formulation}\label{Problem_formulation}

We consider uncharged conducting droplets that are moving within a uniaxial compressional flow $\textit{\textbf{U}}^\infty(\boldsymbol{x}) = (\dot{\gamma}x_1,\dot{\gamma}x_2,-2\dot{\gamma}x_3)$ and subject to a uniform electric field that acts at an angle $\eta$ relative to the compressional axis. Here, $x_1$ and $x_2$ are the two extensional axes, and $x_3$ is the compressional axis. In atmospheric clouds, droplet volume fractions are typically very low (on the order of $10^{-6}$) \citep{grabowski2013growth}, allowing us to disregard interactions involving three or more droplets. Therefore, our analysis focuses on pairwise interactions and collisions. Because the Reynolds number based on the droplet radius is sufficiently small, we can accurately describe the motion of the surrounding fluid phase using the Stokes equations for creeping flow. The negligible inertia of the droplets, combined with the linear nature of Stokes flow, enables us to express the relative velocity between the droplets as a linear superposition of the relative velocities induced by the flow, the external electric field, and van der Waals forces. Consequently, the relative velocity between the droplets, denoted by $\boldsymbol{V}_{12}$, is given by
\begin{eqnarray}
   \boldsymbol{V}_{12}= \boldsymbol{V}_{1} - \boldsymbol{V}_{2} & = & \boldsymbol{E^{\infty}} \boldsymbol{\cdot} \boldsymbol{r}- \Big[A\frac{\boldsymbol{r}\boldsymbol{r}}{r^{2}}+B\left(\boldsymbol{I}-\frac{\boldsymbol{r}\boldsymbol{r}}{r^{2}}\right)\Big] \boldsymbol{\cdot} \left(\boldsymbol{E^{\infty}} \boldsymbol{\cdot} \boldsymbol{r}\right) \nonumber \\ & + & \frac{1}{6\pi\mu_{f}}\left(\frac{1}{a_{1}}+\frac{1}{a_{2}}\right)\left[G\frac{\boldsymbol{r}\boldsymbol{r}}{r^{2}}+H\left(\boldsymbol{I}-\frac{\boldsymbol{r}\boldsymbol{r}}{r^{2}}\right)\right] \boldsymbol{\cdot} \left(\boldsymbol{F}_E + \boldsymbol{F}_{\text{vdW}}\right),
\label{general_relative_velocity_equation} 
\end{eqnarray}
where $\boldsymbol{V}_{1}$ and $\boldsymbol{V}_{2}$ are the velocities of the satellite droplet (radius $a_1$) and the test droplet (radius $a_2$) respectively, $\textit{\textbf{E}}^{\infty} =  [(\boldsymbol{\nabla} \textit{\textbf{U}}^{\infty}) + (\boldsymbol{\nabla} \textit{\textbf{U}}^{\infty})^T]/2$ is the strain rate tensor, $\boldsymbol{r}$ is the vector from the center of the test droplet to the center of the satellite droplet, with $r = |\boldsymbol{r}|$, $\textbf{\textit{I}}$ is the second-order unit tensor, and $\boldsymbol{F}_E$ and $\boldsymbol{F}_{\text{vdW}}$ are the electric-field-induced and van der Waals forces, respectively. The mobility functions $A$, $B$, $G$, and $H$ describe the hydrodynamic interactions, where $A$ and $G$ are axisymmetric mobilities, while $B$ and $H$ are asymmetric mobilities for linear flow and non-hydrodynamic forces, respectively. These functions depend on the droplet size ratio $\kappa = a_2/a_1$ and the dimensionless center-to-center distance $r/a^*$. The methods for computing these mobilities and their asymptotic expressions for small and large separation distances given in \citet{batchelor1972hydrodynamic,batchelor1976brownian,kim1985resistance,jeffrey1992calculation} and \citet{wang1994collision}. In this analysis, we employ the uniformly valid solutions for $A$ and $G$ developed by \citet{dhanasekaran2021collision}, which consider continuum lubrication interactions for $\xi > \textrm{O}(Kn)$ and non-continuum lubrication interactions for $\xi \leq \textrm{O}(Kn)$. Since $B$ and $H$ approach finite values as the separation distance between the two inertia free droplets approach zero, continuum hydrodynamics remain valid for these mobilities at all separation distances.

We adopt a spherical coordinate system $(r,\theta,\phi)$ with the origin at the center of the test droplet. To simplify the analysis, we nondimensionalize the governing equations by choosing $a^*$, $\dot{\gamma}a^*$ and $\dot{\gamma}^{-1}$ as the characteristic length, velocity, and time scales of the problem. Consequently, the nondimensional radial separation between the centers of the two droplets, denoted by $r$ from here onwards, spans the interval from $2$, referred to as the collision sphere, to $\infty$, where one droplet exerts no influence on the other. Similarly, we nondimensionalize the spatial coordinates with $a^*$ and represent these scaled coordinates with an overbar, such that $\overline{x}_1 = x_1/a^*$, $\overline{x}_2 = x_2/a^*$, and $\overline{x}_3 = x_3/a^*$. Here we choose $a_1 > a_2$, permitting the size ratio to vary over the range $(0,1]$.

The electrostatic interaction between two spherical conductors subject to an external electric field is governed by a boundary value problem for Laplace's equation for the electric potential field in a bispherical coordinate system \citep{davis1964two}. Given the resulting potential field, the electric-field-induced forces on each conductor are determined by integrating the electrical stresses over their surfaces. The forces on the two droplets are equal in magnitude, opposite in direction and always act to orient the center-to-center line with the direction of the external electric field. Exploiting the axisymmetric nature of the problem, \citet{davis1964two} expressed these forces by decomposing them into two components: one parallel to the center-to-center line and the other perpendicular to it. The expressions for these forces in the radial ($r$) and angular ($\theta$) directions are given by
\begin{eqnarray}
    F_E^r &=& -4\pi\epsilon_0a_2^2E_0^2\left[F_1\cos^2(\theta-\eta)+F_2\sin^2(\theta-\eta)\right] \qquad \textrm{and}\label{F_E^r} \\ 
    F_E^{\theta} &=& 4\pi\epsilon_0a_2^2E_0^2 F_8\sin 2(\theta-\eta), \label{F_E^theta} 
\end{eqnarray}
where $\theta-\eta$ is the angle of the electric field relative to the center-to-center line of the droplets. The force coefficients $F_1$, $F_2$, and $F_8$ are non-trivial series expressions that depend on the droplet center-to-center distance and the size ratio. The analytical forms of these force coefficients in both the near and far fields are given by \citet{thiruvenkadam2023pair}.

The attractive van der Waals force acts along the center-to-center line of the droplets and is $F_{\text{vdW}}=-d \varPhi_{\text{vdW}}/d r$, where $\varPhi_{\text{vdW}}$ is the van der Waals potential. \citet{hamaker1937london} derived the non-retarded form of $\varPhi_{\text{vdW}}$ using the principle of pairwise additivity. However, retardation effects arise from the frequency dependence of the finite speed of electromagnetic wave propagation (See \citet{RevModPhys.82.1887}, for a review). Retardation effects become significant when the droplet separation distance is comparable to or greater than the London wavelength $\lambda_L (\approx 0.1$ \textmu m). Here, we use the form of retarded van der Waals potential given by \citet{zinchenko1994gravity}, 
$\varPhi_{\text{RvdW}}$, which of course depends on $r$, as well as $\kappa$, the Hamaker constant $A_H$, and the dimensionless parameter $N_L = 2 \pi \left(a_1+a_2\right)/\lambda_L = 2 \pi a_1 \left(1+\kappa\right)/\lambda_L$.

Carrying out the vector and tensor operations in Equation (\ref{general_relative_velocity_equation}) yields the components of the dimensionless relative velocity $\boldsymbol{v}=\boldsymbol{V}_{12}/\dot{\gamma}a^*$ in the $r$, $\theta$, and $\phi$ directions as 
\begin{eqnarray}
    v_r &=& \frac{d r}{d t} = -r\left(1-A\right)\left(3\cos^{2}\theta-1\right) - N_E G \kappa \left(F_1\cos^2(\theta-\eta)  +F_2\sin^2(\theta-\eta)\right) - N_v G \frac{d \varPhi_{\text{RvdW}}}{d r},
    \label{vr_equation} \\
    v_{\theta} &=& r\frac{d \theta}{d t} = 3\left(1-B\right)\sin\theta\cos\theta + N_E H \kappa F_8\sin 2(\theta-\eta) \qquad \textrm{and} \label{vtheta_equation} \\
    v_{\phi} &=& 0. \label{vphi_equation}
\end{eqnarray}
Here, $N_E$ and $N_v$ are dimensionless quantities describing the relative strength of electric-field-induced and retarded van der Waals forces to the background flow and are
\begin{eqnarray}
    N_E &=& \frac{4\epsilon_0 E_0^2}{3\mu_f\dot{\gamma}} \qquad \textrm{and}
    \label{expression_for_NE} \\
    N_v &=& \frac{2 A_H}{3\pi\mu_f\dot{\gamma}\kappa\left(1+\kappa\right)a_1^3}, \label{expression_for_NF} 
\end{eqnarray}
respectively. The parameter $N_E$ is the principal indicator of how the strength of the electric field influences the collision efficiency, and we have deliberately excluded the size ratio term from its definition to facilitate a clear understanding of the interrelationship between $N_E$ and the collision dynamics. However, consistent with the definitions given in preceding studies that explored the effects of van der Waals forces on droplet interactions within a linear flow, the parameter $N_v$ does depend upon the size ratio. In clouds, depending on the strength of the electric field and compression rate, the value of $N_E$ can vary from O$(10^{-4})$ to O$(10^4)$. Similarly, depending on the sizes of droplets $N_v$ varies from O$(10^{-5})$ to O$(10^0)$.

The collision rate, $K_{12}$, quantifies the rate at which two species of given number densities collide with each other per unit volume. Mathematically, $K_{12}$ is equivalent to the flux of droplets into the collision surface and can be expressed as follows, 
\begin{equation}
   K_{12} = -n_1n_2 \dot{\gamma} \left(a^*\right)^3 \int_{(r=2)\&\left(\boldsymbol{v}\boldsymbol{\cdot}\boldsymbol{n}<0\right)}  \left(\boldsymbol{v}\mathbf{\cdot}\textbf{\textit{n}}\right) P(r) dA,
\label{Collision_rate_general_dimensional}
\end{equation}
where $P(r)$ is the pair distribution function and $\boldsymbol{n}$ denotes the outward unit normal at the collision surface. In this context, the condition $\boldsymbol{v}\boldsymbol{\cdot}\boldsymbol{n}<0$ in Eq. (\ref{Collision_rate_general_dimensional}) indicates that the droplet radial velocity must be directed inwards at the collision surface for two droplets to collide. In a dilute system, such as in clouds, the pair distribution function satisfies the quasi-steady Fokker-Planck equation:
\begin{equation}
    \boldsymbol{\nabla}\boldsymbol{\cdot}\left(P\boldsymbol{v}\right) = 0.
\label{Probability_conservation}
\end{equation}
In the far field, droplet motions become uncorrelated, so that the condition $P \rightarrow 1$ as $r \rightarrow \infty$ is satisfied. For the purposes of calculation, we consider $r=r_{\infty}$, which is very large but finite.

The absence of diffusive flux in Eq. (\ref{Probability_conservation}) allows us to determine the collision rate through trajectory analysis. We use Eq. \eqref{Probability_conservation} and the divergence theorem to evaluate the integral in Eq. (\ref{Collision_rate_general_dimensional}) over the surface that encloses the volume occupied by all trajectories starting at $r=r_{\infty}$ and ending at $r=2$. Consequently, the flux through the cross-section of this volume at $r=r_{\infty}$ defines the collision rate. We define this cross-section as the upstream interception area, $A_c$. Now, at $r=r_{\infty}$, non-hydrodynamic forces become negligible, so that the background flow is the sole driving force underlying the relative velocity between the droplets. Because the pair distribution function is $P=1$ at $r=r_{\infty}$, we can simplify Eq. (\ref{Collision_rate_general_dimensional}) as follows
\begin{equation}
    K_{12} = -n_1n_2 \dot{\gamma} \left(a^*\right)^3\int_{A_c}\left(\boldsymbol{v}\boldsymbol{\cdot}\boldsymbol{n}'\right)|_{r_{\infty}} dA,
\label{Collision_rate_upstream_interception_area}
\end{equation}
where $\boldsymbol{n}'$ is the outward normal unit vector on the area element of $A_c$.

In the absence of hydrodynamic interactions and non-hydrodynamic forces, the trajectories of inertialess droplets coincide with the undisturbed streamlines of the background flow. Therefore, in this scenario, $P=1$ for all droplet separation distances. The collision rate obtained without interactions is called the ideal collision rate and is denoted as $K_{12}^{0}$. \citet{zeichner1977use} determined the ideal collision rate for a uniaxial compressional flow as
\begin{equation}
    K^{0}_{12} = \frac{8\pi}{3\sqrt{3}}n_1n_2\dot{\gamma}\left(a_{1}+a_{2}\right)^{3}.
\label{ideal_collision_rate}
\end{equation}
The collision efficiency, $E_{12}$, is defined as the ratio of actual collision rate, $K_{12}$, to the ideal collision rate, $K_{12}^{0}$, viz.
\begin{equation}
    E_{12} = \frac{K_{12}}{K_{12}^{0}}.
\label{Collision_efficiency}
\end{equation}

It is clear from Eq. (\ref{Collision_rate_upstream_interception_area}) that calculating the collision rate requires determining the upstream interception area, which we do using a trajectory analysis methodology. This involves initializing a test droplet at the origin and evolving the satellite droplets to identify those that collide, thereby defining $A_c$.

Eqs. \eqref{vr_equation} and \eqref{vtheta_equation} give the components of the relative velocity between a pair of hydrodynamically interacting droplets, influenced by the combined effects of a uniaxial compressional flow, an external electric field, and van der Waals forces.  Combining these components allows us to compute the relative trajectories of the droplets by integrating the following dimensionless trajectory equation:
\begin{equation}
    \frac{d\theta}{dr}=\frac{1}{r}\frac{v_{\theta}}{v_r}.
    \label{final_trajectory_equation}
\end{equation}
The colliding trajectories correspond to the paths traversed by the centers of evolving satellite droplets, which start from a position far upstream and end at the collision surface. Analyzing these colliding trajectories in the far field is crucial for determining the upstream interception area. However, if we choose initial conditions over the spherical shell at $r_{\infty}$, the computational burden associated with trajectory calculations can be considerable,  since most of the trajectories initiated from this shell do not reach the collision sphere. To circumvent this complication, we exploit the quasi-steady nature of the trajectory equation and consider time-reversed trajectories initialized on the collision surface. This methodological adjustment substantially reduces the required number of trajectories for computation. To further streamline the selection of trajectory starting points, we restrict our focus to those positions on the collision sphere where the radial velocity is negative; $v_r < 0$. It is important to note that at $r = 2$ the hydrodynamic mobilities are $A=1$ and $G=0$, so that $v_r = 0$. To avoid this issue, we set initial conditions on a sphere of radius $r=2+\delta$, where $\delta$ is a small offset distance from the collision surface. We demonstrate that converged results are achievable with minimal computational effort when $\delta = 10^{-6}$. With these carefully chosen initial conditions, we conduct backward integrations of equation (\ref{final_trajectory_equation}) using a fourth-order Runge-Kutta method. We note that, depending on the computational requirements, we may alternatively solve for $r(\theta)$ from $dr/d\theta=r v_r/v_{\theta}$ rather than solving for $\theta(r)$ from Eq. (\ref{final_trajectory_equation}).

\section{Results and discussion}\label{Results_and_discussion}

Our primary aim is to quantitatively assess the impact of an external electric field on the relative trajectories and collision efficiency of two conducting droplets in a compressional airflow. To establish a clear context for this investigation, we first briefly review the pair trajectories associated with two distinct scenarios: one where a uniaxial compressional flow alone ($N_E=0$) and the other where an external electric field alone ($N_E=\infty$) drives the relative motion between two hydrodynamically interacting droplets. We then analyze pair trajectories for the combined effects of the flow and electric field, that is for finite $N_E$. In this problem, the relative velocity between the droplet pair is independent of the azimuthal coordinate $\phi$, and hence we examine the relative trajectories in a representative $r\sin\theta - r\cos\theta$ plane.

\begin{figure}
\centering
\includegraphics[width=1.0\textwidth]{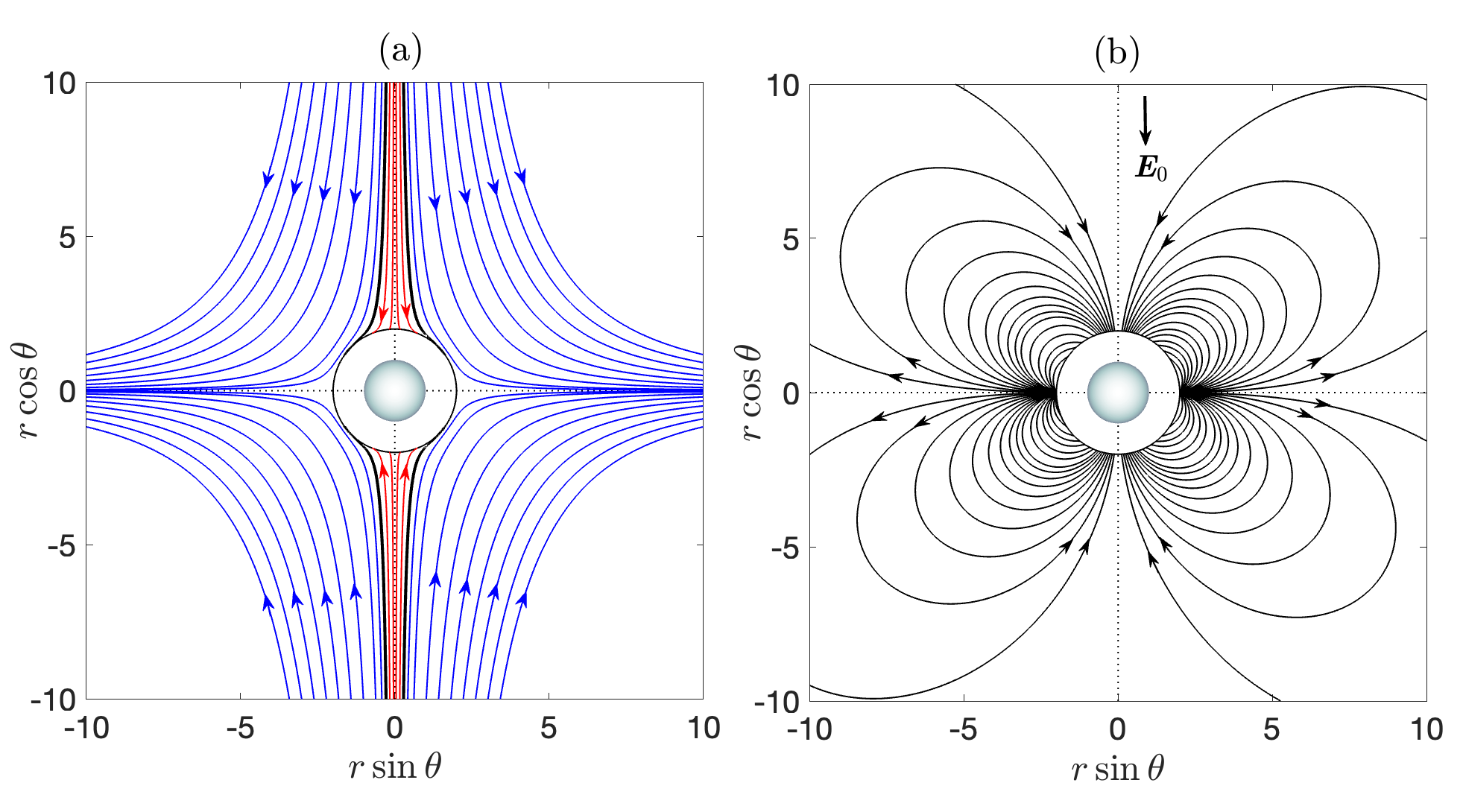}
\caption{Relative trajectories of two equal-sized hydrodynamically interacting droplets in (a) a uniaxial compressional flow for $Kn=10^{-2}$ and (b) a vertical electric field. The blue, red, and black lines denote open, colliding, and loop trajectories respectively. The sphere at the center represents the test droplet, and the thin black circle represents the projection of the colliding sphere. The arrows on the trajectories show representative trajectory directions.}
\label{Pair_trajectories_zero_electric_field_and_only_electric_field}
\end{figure}

\begin{figure}
\centering
\includegraphics[width=1.0\textwidth]{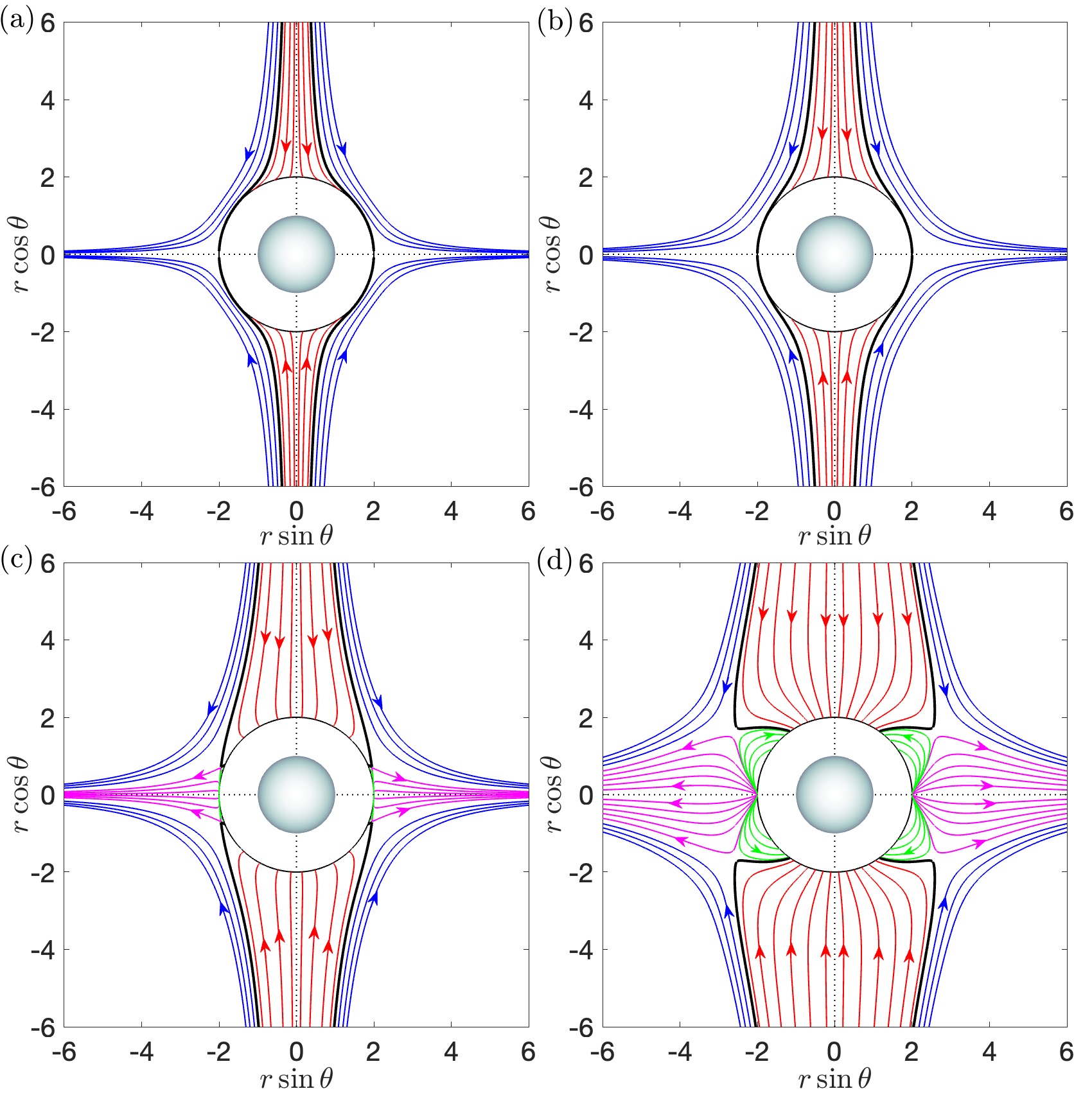}
\caption{Pair trajectories of two inertialess uncharged conducting droplets subject to a uniaxial compressional flow and a vertical electric field ($\eta=0$) when $\kappa=0.5$, $Kn=10^{-2}$, $N_v=10^{-3}$, and (a) $N_E=2 \times 10^{-1}$, (b) $N_E=2$, (c) $N_E=20$, and (d) $N_E=2 \times 10^2$. The blue, green, red, and thick black lines are open, loop, colliding, and limiting colliding trajectories. Pink lines are a separate class of trajectory that start from two specific locations on the collision surface and diverge to infinity.}
\label{Pair_trajectories_electric_field_alpha_0}
\end{figure}

\begin{figure}
\centering
\includegraphics[width=1.0\textwidth]{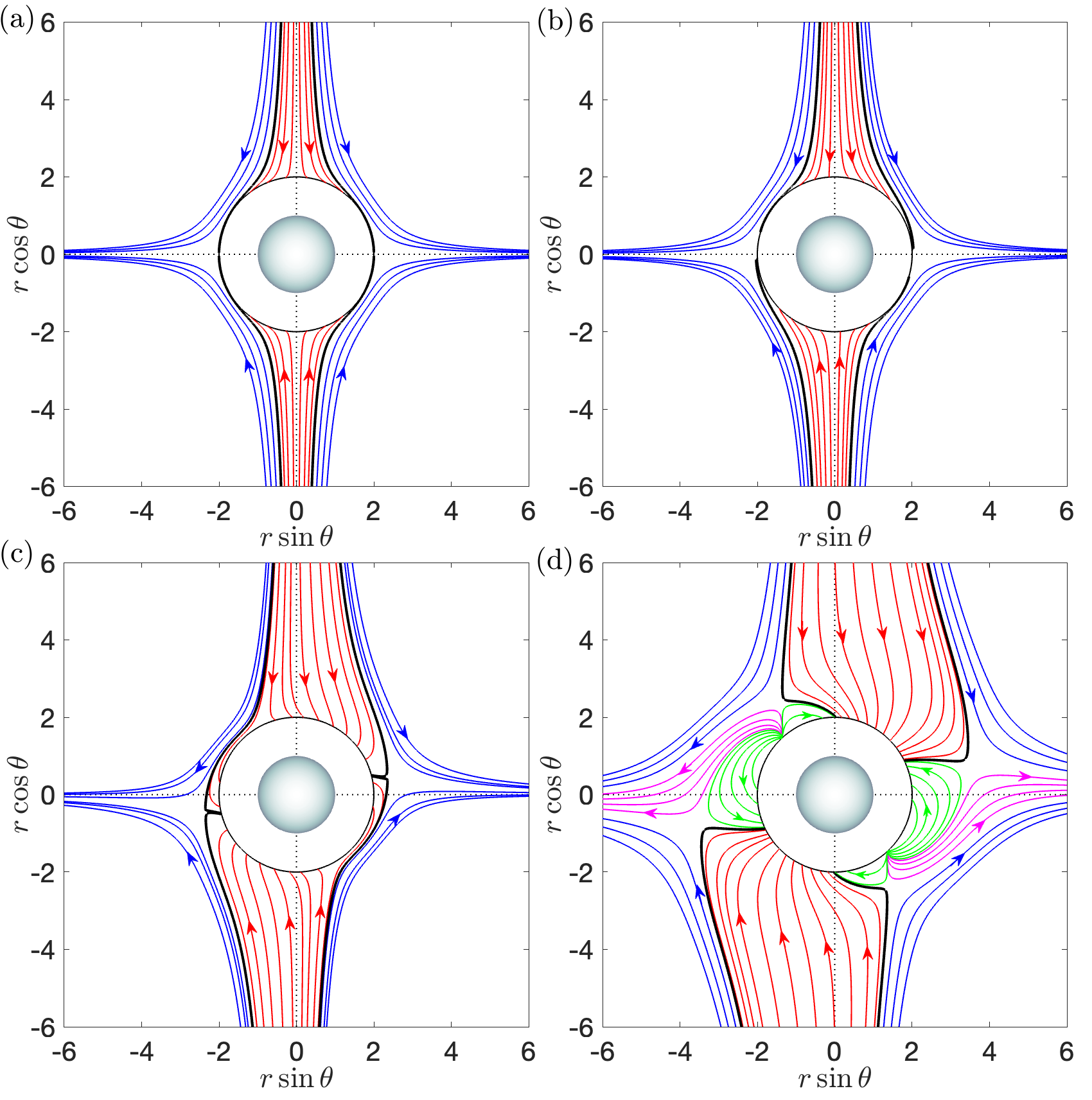}
\caption{Same as figure \ref{Pair_trajectories_electric_field_alpha_0} except that $\eta=\pi/4$}
\label{Pair_trajectories_electric_field_alpha_45}
\end{figure}

\begin{figure}
\centering
\includegraphics[width=1.0\textwidth]{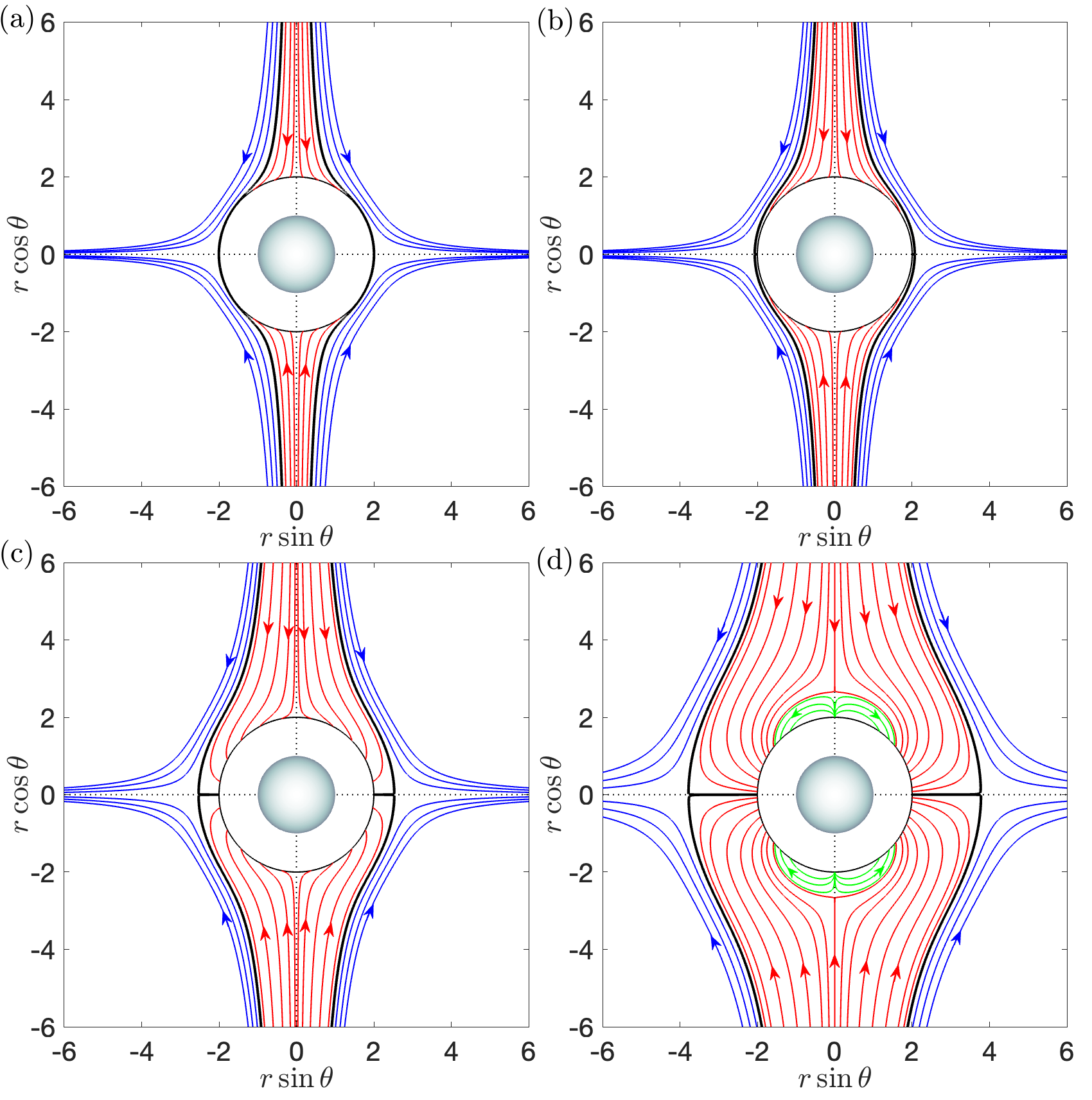}
\caption{Same as figure \ref{Pair_trajectories_electric_field_alpha_0} except that $\eta=\pi/2$}
\label{Pair_trajectories_electric_field_alpha_90}
\end{figure}

Figure \ref{Pair_trajectories_zero_electric_field_and_only_electric_field}(a) shows the pattern of relative trajectories in a uniaxial compressional flow, where the vertical axis corresponds to the axis of compression. With non-continuum hydrodynamics, there are two distinct types of relative trajectories: (i) open trajectories that arrive from infinity and depart to infinity without reaching the collision sphere (blue lines), and (ii) colliding trajectories that arrive from infinity and reach the collision sphere (red lines). These trajectories are fore–aft symmetric. The analytical expression for these relative trajectories, derived by integrating Eq. (\ref{final_trajectory_equation}) subject to $N_E=N_v=0$, is given by
\begin{equation}
    \sin^2\theta (r)\cos\theta (r) = C \varphi (r) , \label{relative_trajectory_expression_only_flow}
\end{equation}
where $C$ is the constant specifying a particular trajectory and
\begin{equation}
    \varphi (r) = \exp\left[\int_2^r\frac{3(B-1)}{r(1-A)}dr\right]. \label{varphi(r)_expression} 
\end{equation}
By examining the sign of the radial relative velocity at the collision sphere, it is straightforward to establish that open and colliding trajectories correspond to $|C| > 2/(3\sqrt{3})$ and $|C| < 2/(3\sqrt{3})$, respectively. The limiting colliding trajectories for which $|C| = 2/(3\sqrt{3})$ touch the collision sphere at one of the four locations: $\theta=\theta_c, \pi-\theta_c, \pi+\theta_c, 2\pi-\theta_c$ where $\theta_c = \arccos(1/\sqrt{3})$. These limiting colliding trajectories act as separatrices between open and colliding trajectories. Here, they form the boundaries of the two upstream interception areas, which are circles centered at two extremes of the compressional axis. Finally, the analytical expression for the collision efficiency, derived in terms of relevant hydrodynamic mobilities, is given by \citet{wang1994collision} as 
\begin{equation}
    E_{12} = \exp\left[\int_2^{\infty}\frac{3(B-A)}{r(1-A)}dr\right]. \label{Analytical_expression_collision_efficiency_only_flow} 
\end{equation}

In contrast, when the electric field alone dictates the dynamics, the relative trajectories begin and end on the collision sphere, forming what we call loop trajectories. Figure \ref{Pair_trajectories_zero_electric_field_and_only_electric_field}(b) shows a map of these loop trajectories for a vertical electric field. In this two-dimensional dynamical system, $\theta=\pi/2$ and $3\pi/2$ are two unstable fixed points on the collision sphere $r=2$. Droplet centers start near these two points and follow trajectories either in the first and fourth or second and third quadrants. A detailed discussion of relative trajectories in an electric field is given by \citet{thiruvenkadam2023pair}.

A comprehensive investigation of the evolution of relative trajectory topologies under varying strengths of the electric field and compressional flow is essential for an accurate description of collision dynamics. Figures \ref{Pair_trajectories_electric_field_alpha_0}, \ref{Pair_trajectories_electric_field_alpha_45}, and \ref{Pair_trajectories_electric_field_alpha_90} show typical pair trajectories for three different angles between the electric field and the compressional axis: $\eta = 0$, $\eta = \pi/4$, and $\eta = \pi/2$, respectively.  For each figure we fix $\kappa = 0.5$, $Kn = 10^{-2}$ and $N_v = 0$, and show four panels with (a) $N_E = 2 \times 10^{-1}$, (b) $N_E = 2$, (c) $N_E = 20$, and (d) $N_E = 2 \times 10^2$.

When the electric field is relatively small, such as for $N_E = 2 \times 10^{-1}$ and $N_E = 2$, the pair trajectories resemble those observed in the absence of an electric field, where only open and colliding trajectories are present. As we increase $N_E$, colliding trajectories tend to converge toward impact locations centered around $\theta = \eta$ and $\theta = \eta + \pi$, and loop trajectories begin to emerge. Depending on the value of $\eta$, these loop trajectories initiate at $r = 2, \theta = \eta + (\pi/2)$ and $r = 2, \theta = \eta + (3\pi/2)$, and terminate at various locations on the collision sphere. As previously discussed, these starting locations are unstable fixed points. However, not all trajectories that originate from these locations lead to loops. Notably, some trajectories (depicted by pink lines) diverge to infinity instead, and the volume occupied by these trajectories increases as $N_E$ increases.

When the electric field is relatively large, such as for $N_E=2 \times 10^2$, the forces induced by the electric field dominate the dynamics at small droplet separations. This results in near field trajectories resembling the loop trajectories that would exist in the absence of flow. Furthermore, colliding trajectories bend sharply at intermediate separation distances before following the loops and eventually impacting the collision sphere. The influence of the electric field and the background linear flow becomes comparable at these bending points. The flow dictates the relative motion for larger droplet separation distances, causing far-field trajectories to converge to those determined by flow alone. The limiting colliding trajectories, shown as thick black lines in figures \ref{Pair_trajectories_electric_field_alpha_0}, \ref{Pair_trajectories_electric_field_alpha_45}, and \ref{Pair_trajectories_electric_field_alpha_90}, align with open trajectories in the far field and loop trajectories in the near field, consistent with the expected behaviors in both fields. Except for $\eta=\pi/2$, these limiting trajectories act as separatrices, distinguishing loop and colliding trajectories in the near field and open and colliding trajectories in the far field. Although not explicitly illustrated in the figures, for sufficiently large $N_E$, the outermost pink trajectories and limiting colliding trajectories intersect at points of maximum curvature for both curves. These intersections define saddle points, whose locations depend strongly on the value of $\eta$.

The preceding trajectory analysis showed that the strength of the electric field and its inclination angle relative to the compressional axis significantly influence the radii and locations of the centers of the circular upstream interception areas. For a given $\eta$, the size of the upstream area increases with the electric field strength (i.e., as $N_E$ increases). This is due to the expanding sphere of influence of the attractive force induced by the electric field as its strength increases. Notably, except at asymptotically small values of $N_E$ and for $\eta=0,\pi/2$, the electric field breaks the fore-aft symmetry of the trajectories, resulting in a shift of the centers of the two upstream areas away from the compressional axis.

\begin{figure}
\centering
\includegraphics[width=1.0\textwidth]{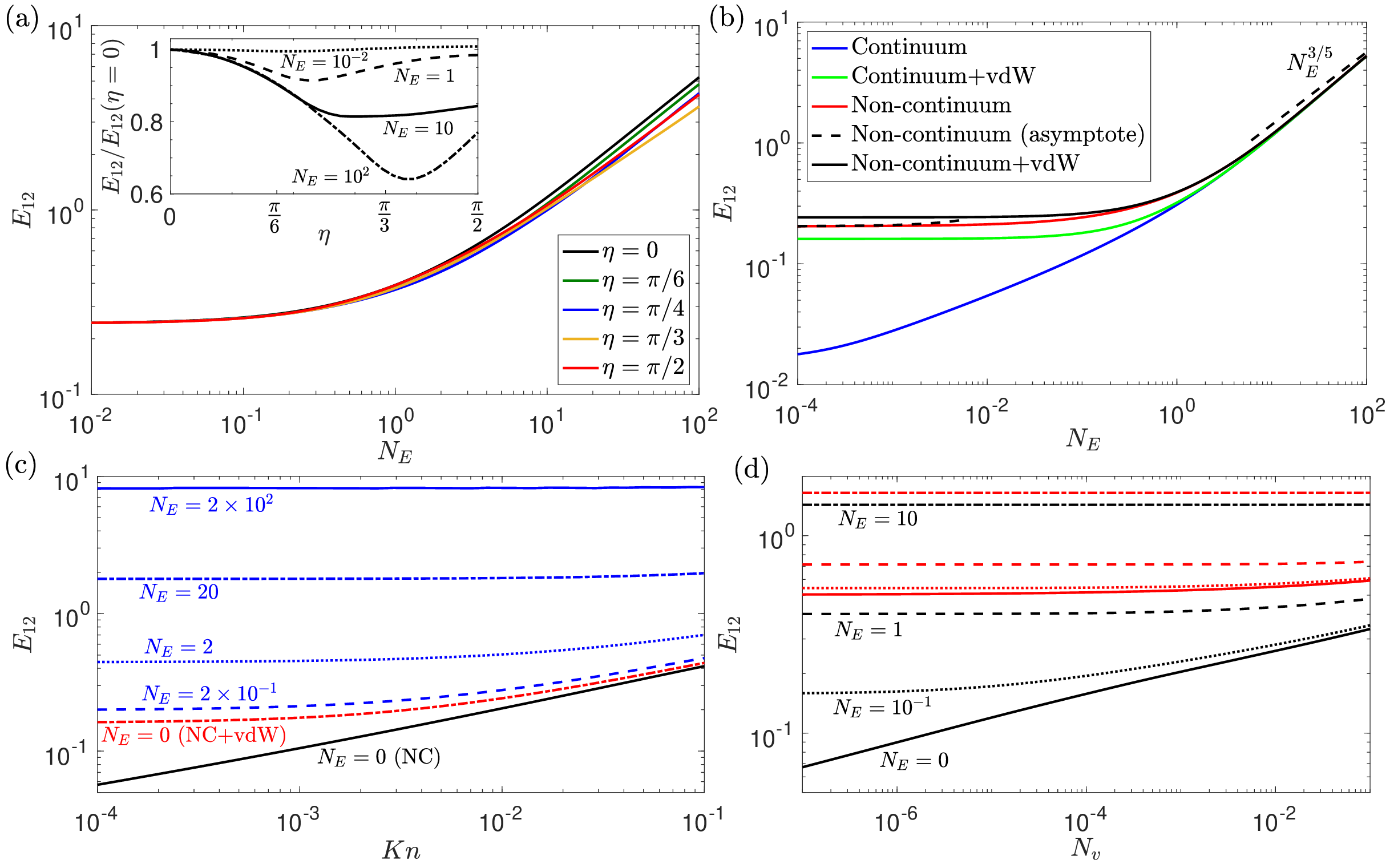}
\caption{(a) Variation of collision efficiency with the relative strength of electrostatic effects and compressional flow, characterized by $N_E$, for $\eta=0, \pi/6, \pi/4, \pi/3$ and $\pi/2$ when $\kappa = 0.5$, $Kn = 10^{-2}$ and $N_v = 0$. The inset shows how collision efficiency scaled by its value at $\eta=0$ varies with $\eta$ for the same values of $\kappa$, $Kn$, and $Nv$ when $N_E = 10^{-2}, 1, 10$ and $ 10^2$. (b) Variation of collision efficiency with $N_E$ for different collision-inducing mechanisms when $\eta=0$. The black dashed line shows the asymptotic behavior for non-continuum (NC) lubrication effects. All parameters are the same as in (a) except for the cases with der Waals force where $N_L = 500$ and $N_v = 10^{-3}$. (c) Collision efficiency as a function of $Kn$ for non-continuum hydrodynamics plus van der Waals and electric-field-induced forces with $N_E=2\times 10^{-1}, 2, 20$ and $2\times 10^2$ when $\kappa = 0.5$, $N_L = 250$, $N_v = 10^{-3}$ and $\eta=0$. Continuous black lines and dash-dotted red lines are for the case without an electric field. (d) Collision efficiency as a function of $N_v$ for continuum (black lines) and non-continuum ($Kn=10^{-1}$, red lines) lubrication interactions when $\kappa = 0.9$, $N_L = 500$, $\eta=0$ and $N_E = 0, 10^{-1}, 1, 10$.}
\label{Collision_efficiency_with_NE_Kn_and_Nv}
\end{figure}

Having shown how the trajectory analysis underlies how the upstream interception area influences the collision efficiency, we now
examine how the collision efficiency depends on the key physical quantities involved in this problem. Figure \ref{Collision_efficiency_with_NE_Kn_and_Nv}(a) illustrates how the collision efficiency varies with the relative strength of the electric-field-induced force and background flow for five different orientations of the electric field ($\eta=0, \pi/6, \pi/4, \pi/3,\pi/2$) when we have $\kappa = 0.5$, $Kn = 10^{-2}$, and $N_v = 0$. In the absence of van der Waals forces, non-continuum hydrodynamics drives the collision mechanism in the flow-dominated regime (i.e., $N_E \ll 1$). Since the droplet pair experiences the same non-continuum lubrication effects for a given $\kappa$ and $Kn$, the curves corresponding to different $\eta$ tend to converge in the small $N_E$ regime before they asymptote to the value that corresponds to the collision efficiency of the droplet pair in a uniaxial compressional flow with non-continuum lubrication interactions. For fixed $\eta$, the collision efficiency increases monotonically with increasing $N_E$ and exhibits  power-law growth when $N_E \gg 1$, the exponent of which varies between $0.6 - 0.65$ depending on the specific values of $\kappa$ and $\eta$.  To further examine the dependency of collision efficiency on $\eta$, we compute $E_{12}$ for fixed values of $N_E$ while varying $\eta$ from $0$ to $\pi/2$. The inset in Figure \ref{Collision_efficiency_with_NE_Kn_and_Nv}(a) shows the normalized collision efficiency, scaled by its value at $\eta = 0$, as a function of $\eta$ for $N_E = 10^{-2}, 1, 10,$ and $10^2$, and the same values of $\kappa$, $Kn$ and $N_v$. For moderate to large values of $N_E$, as $\eta$ increases, the scaled collision efficiency initially decreases to a minimum before subsequently rising. As $N_E$ increases, this minimum shifts towards higher values of $\eta$. Conversely, for a small value of $N_E$ (e.g., $N_E=10^{-2}$), the collision efficiency remains almost constant with respect to increases in $\eta$. This behavior is indicative of a slight perturbative effect induced by a weak electric field on the collision efficiency of two droplets in a uniaxial compressional flow with non-continuum lubrication interactions. To capture these perturbation effects quantitatively, in the Appendix we derive an analytical expression for collision efficiency up to order $N_E$.

To compare the importance of various collision-inducing mechanisms, in figure \ref{Collision_efficiency_with_NE_Kn_and_Nv}(b), we show the collision efficiency as a function of $N_E$ for droplets interacting through continuum and non-continuum hydrodynamics, with or without van der Waals forces. The collision efficiency decreases rapidly as $N_E$ decreases when the droplet pair interacts via full continuum hydrodynamics (blue line in figure \ref{Collision_efficiency_with_NE_Kn_and_Nv}(b)). In this case, due to the absence of collision-inducing mechanisms, $E_{12}$ will approach zero as $N_E \rightarrow 0$. However, attractive van der Waals forces can cause collisions by overcoming the continuum lubrication resistance in the flow-dominated regime. Thus, the collision efficiency with continuum hydrodynamics plus van der Waals interactions asymptotes to a finite value as $N_E \rightarrow 0$ (green line in figure \ref{Collision_efficiency_with_NE_Kn_and_Nv}(b)). As expected, for small to moderate values of $N_E$, the collision efficiency resulting from non-continuum hydrodynamics augmented by van der Waals interactions (shown by the black line in figure \ref{Collision_efficiency_with_NE_Kn_and_Nv}(b)) surpasses that arising from non-continuum hydrodynamics alone. The asymptote shown by the black dashed line demonstrates the validity of the analytical expression for the collision efficiency up to O$(N_E)$. In regimes characterized by strong electric fields, where electrostatic forces dominate, collision efficiencies converge for all evaluated scenarios. In this specific example, where $\kappa=0.5$, $Kn=10^{-2}$ and $\eta=0$, in the electric-field-dominated regime the collision efficiency scales as $N_E^{3/5}$, which is a consequence of the following simple argument.  
When the first and second terms in the radial relative velocity equation (\ref{vr_equation}) are comparable, the droplet center-to-center distance
$r_\textrm{crit}$ is large. Now, in the far field $A \rightarrow 0$, $G \rightarrow 1$, $F_1 \sim 96\lambda(1-\lambda)^3/r^4$, and $F_2 \sim -48\lambda(1-\lambda)^3/r^4$, where $\lambda=\kappa/(1-\kappa)$. Hence, by balancing the radial relative velocity due to the linear flow with that due to the electric field, we find that when $\eta=0$, we have $r_\textrm{crit} \propto N_E^{1/5}$. Therefore, the collision efficiency $E_{12} \sim r_\textrm{crit}^3 \propto N_E^{3/5}$, as shown in figure \ref{Collision_efficiency_with_NE_Kn_and_Nv}(b). 


Figure \ref{Collision_efficiency_with_NE_Kn_and_Nv}(c) illustrates how collision efficiency varies with the strength of non-continuum lubrication effects, as measured by the Knudsen number $Kn$, for $N_E = 2 \times 10^{-1}, 2, 20, 2 \times 10^2$, $\kappa=0.5$, $N_L=500$, $N_v=10^{-3}$, and $\eta=0$. To emphasize the impact of the electric field, we compare our results with cases that do not account for electrostatic forces. The collision efficiency resulting from non-continuum lubrication effects (NC) alone is shown by the black line in \ref{Collision_efficiency_with_NE_Kn_and_Nv}(c), calculated by evaluating the integral in Eq. (\ref{Analytical_expression_collision_efficiency_only_flow}) for a range of $Kn$. As $Kn$ decreases, so too does the relative thickness of the non-continuum lubrication layer, leading to a monotonic decrease in $E_{12}$ due to NC alone, which approaches zero in the limit as $Kn \rightarrow 0$. Incorporating van der Waals (vdW) interactions modifies the asymptotic behavior of the collision efficiency in the small $Kn$ regime, where non-continuum effects become negligible (as shown by the red dash-dotted line in figure \ref{Collision_efficiency_with_NE_Kn_and_Nv}(c)). Consequently, as $Kn$ decreases, the collision efficiency for the NC+vdW case decreases and ultimately asymptotes to a value representative of the collision efficiency controlled by van der Waals forces for a droplet pair interacting through continuum hydrodynamics in a uniaxial compressional flow. Including electric-field-induced forces in the NC+vdW case predictably enhances collision efficiency. The dependence of $E_{12}$ on $Kn$ qualitatively resembles the NC+vdW case for weak electric fields (e.g., $N_E = 2\times 10^{-1}, 2$). However, in a strong electric field, the electric-field-induced forces are stronger than the combined effects of NC and vdW interactions, so that the collision efficiency becomes independent of $Kn$. The lines for $N_E = 20$ and $N_E=2 \times 10^2$ in figure \ref{Collision_efficiency_with_NE_Kn_and_Nv}(c) illustrate these scenarios.

To explore the influence of van der Waals forces on collision dynamics, we compute the collision efficiency from weak ($N_v=10^{-7}$) to strong ($N_v=10^{-1}$) van der Waals interactions. In Figure \ref{Collision_efficiency_with_NE_Kn_and_Nv}(d), we illustrate how collision efficiency varies with $N_v$ when $\kappa = 0.9$, $N_L = 500$, and $N_E = 0, 10^{-1}, 1,$ and $10$. The red lines show results with non-continuum lubrication effects, denoted by a Knudsen number of $Kn=10^{-1}$, and the black lines show results using continuum hydrodynamic interactions. Our analysis reveals that for a given $N_E$, the values of $E_{12}$ with non-continuum conditions consistently exceed those with continuum hydrodynamics. As anticipated, the collision efficiency decreases as $N_v$ decreases, which is particularly noticeable in the continuum hydrodynamic cases with either no electric field ($N_E=0$) or weak electric fields ($N_E=10^{-2}$). In contrast, with non-continuum lubrication interactions, the decrease in $E_{12}$ is more gradual, ultimately approaching a value congruent with the collision efficiency calculated without van der Waals interactions. Notably, when the imposed electric field strength is large, such as for $N_E = 10$, the collision efficiency exhibits a remarkable independence to variations in the van der Waals force, implying that electric-field-induced forces dominate the other collision-inducing mechanisms.

\begin{figure}
\centering
\includegraphics[width=1.0\textwidth]{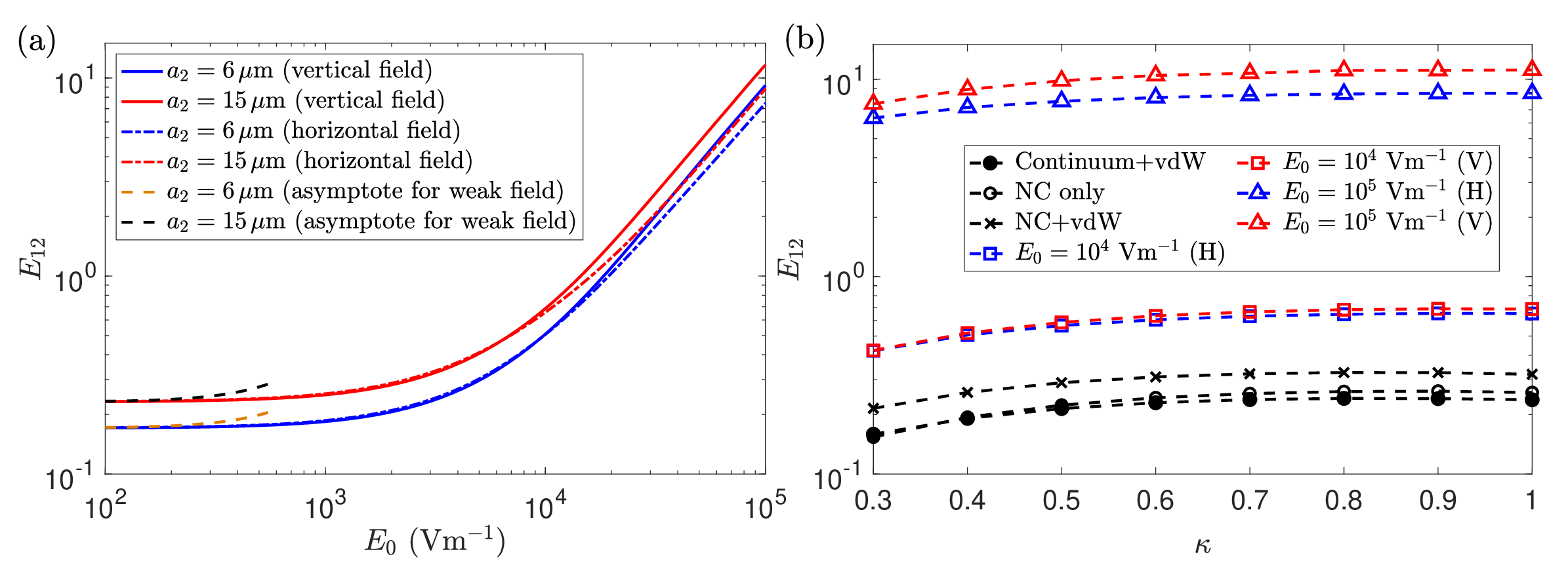}
\caption{(a) Collision efficiency as a function of the strength of the vertical ($\eta=0$) or horizontal ($\eta=\pi/2$) electric field for pairs of droplets with sizes $a_1=15$ \textmu m and $a_2 = 6, 15$ \textmu m, assuming the absence of van der Waals forces. Continuous and dash-dotted lines represent vertical and horizontal electric fields, respectively. Dashed lines are from the asymptotic expression of collision efficiency for a weak electric field given by Eq. (\ref{Analytical_expression_weak_field}). (b) Collision efficiency as a function of the size ratio for water droplets in air with $a_1 = 10$ \textmu m, vertical (indicated by ``V" within parentheses) and horizontal (indicated by ``H" within parentheses) electric fields of $10^4$ Vm$^{-1}$ and $10^5$ Vm$^{-1}$ when non-continuum effects (NC), van der Waals interactions (vdW) and electric-field-induced forces act together.To compare these findings, we present results from \citet{dhanasekaran2021collision}, in which the authors predicted collision efficiencies without an external electric field. The three black lines delineate the efficiencies corresponding to $E_0 = 0$.}
\label{Collision_efficiency_with_E0_and_kappa}
\end{figure}

Having examined the influence of key parameters on droplet collision dynamics, we now focus on the dependence of collision efficiency on the electric field strength ($E_0$) observed in clouds. We take the compression rate of $\dot{\gamma}=25$ s$^{-1}$ and consider two distinct sets of droplet pairs: one with a significant size disparity of $a_1 = 15$ \textmu m and $a_2 = 6$ \textmu m ($\kappa = 0.4$), and the other with equal sized droplets of $a_1 = a_2 = 15$ \textmu m ($\kappa = 1.0$). (Note that in Sec. \ref{Introduction} we gave the values of the densities of water droplets $\rho_p$, and air $\rho_f$, and the dynamic viscosity of air $\mu_f$, in typical clouds.) The mean free path of air tends to increase with altitude in the troposphere, and its value for warm clouds is approximately $0.1$ \textmu m (see \citet{wallace2006atmospheric}). Therefore, the expression for the Knudsen number as a function of the size ratio becomes $Kn = 0.013/(1+\kappa)$. The Hamaker constant for water droplets in air is approximately $3.7 \times 10^{-20}$ J (see \citet{friendlander2000smoke}). Accordingly, the dependencies of $N_L$ and $N_v$ on $\kappa$ are $N_L = 9.42 \times 10^2 (1+\kappa)$ and $N_v = 5.5 \times 10^{-3}/[\kappa (1+\kappa)]$, so that $N_E$ varies with the strength of the electric field as $N_E= 2.77 \times 10^{-8} E_0^2$. Using these relationships, we obtain the values of $Kn$, $N_L$, $N_v$, and $N_E$ that are essential for calculating the upstream interception area and the collision efficiency. As expected and shown in figure \ref{Collision_efficiency_with_E0_and_kappa}(a), the dependence of the collision efficiency on the strength of a vertical or horizontal electric field is qualitatively similar to that of $E_{12}$ on $N_E$ that we showed in figure \ref{Collision_efficiency_with_NE_Kn_and_Nv}(a). For electric field magnitudes $E_0$, reaching a few thousand Vm$^{-1}$, there is only an incremental increase in $E_{12}$, indicating a negligible contribution of the fair-weather electric field on droplet collisions. The dashed lines in figure \ref{Collision_efficiency_with_E0_and_kappa}(a) show results derived from the analytical form of $E_{12}$ in Eq. (\ref{Analytical_expression_weak_field}) for weak electric fields, demonstrating the accuracy of the
asymptotic prediction of collision efficiency influenced by the fair-weather electric field. As the electric field strength exceeds $10^4$ Vm$^{-1}$, which is characteristic of strongly electrified clouds, the collision efficiency is enhanced, as shown in figure \ref{Collision_efficiency_with_E0_and_kappa}(a). In the case of strong vertical electric fields, since $N_E \propto E_0^2$, we have $E_{12} \sim E_0^{6/5}$. These findings show that a vertical electric field exerts a more pronounced effect than does a horizontal electric field in promoting collisions between droplets in a uniaxial compressional flow.

Finally, we examine the influence of droplet polydispersity on collision dynamics when non-continuum hydrodynamics, van der Waals forces, and electric-field-induced forces are operative. Figure \ref{Collision_efficiency_with_E0_and_kappa}(b) shows the dependence of collision efficiency on the droplet size ratio for vertical or horizontal electric field strengths of $10^4$ Vm$^{-1}$ and $10^5$ Vm$^{-1}$, when $a_1 = 10$ \textmu m and $\dot{\gamma}=25$ s$^{-1}$. Here, the size ratio dependencies are $Kn = 0.02/(1+\kappa)$, $N_L = 6.28 \times 10^2 (1+\kappa)$ and $N_v = 1.8 \times 10^{-2}/[\kappa (1+\kappa)]$. Since $N_E$ does not depend on droplet size, its dependence on $E_0$ discussed above is the same. Collision efficiencies are weaker for droplets with smaller size ratios. This is because when the size difference between two interacting droplets is large, the smaller droplet tends to follow the flow streamlines and moves around the larger droplet without colliding. Therefore, in the flow-dominated regime, droplets can collide if the smaller droplet follows a streamline very close to the larger droplet. Moreover, weakening electric-field-induced forces in conjunction with decreasing size ratios further diminishes collision efficiency. Our findings indicate that for a given droplet pair, the collision efficiency increases by an order of magnitude as the strength of the vertical or horizontal electric field increases from $10^4$ Vm$^{-1}$ to $10^5$ Vm$^{-1}$. As discussed in \S \ref{Introduction}, collisions of spheres in a uniaxial compressional flow with continuum and non-continuum hydrodynamics and van der Waals forces have been studied extensively \citep{wang1994collision,dhanasekaran2021collision}. When comparing these results, shown by the three black lines in figure \ref{Collision_efficiency_with_E0_and_kappa}(b), with the influence of an external electric field shows that the latter invariably enhances collision efficiency.

\section{Conclusions}\label{Conclusions}

Motivated by the microphysics of clouds, we have quantified the influence of an external electric field on the collision dynamics of uncharged water droplets within a uniaxial compressional air flow. We have captured the complex interplay between the multiple forces that control collision efficiency and find that at close separations, electric-field-induced forces and non-continuum hydrodynamic effects can overcome lubrication resistance, facilitating surface-to-surface contact in finite time. By mapping typical pair trajectories, we have provided a comprehensive framework for calculating the upstream interception area, which is a vital quantity for determining collision efficiency. The findings establish that the collective influence of electric-field-induced forces and non-continuum hydrodynamic effects significantly enhance droplet collisions.

While our analysis offers valuable insights into the dynamics of droplet collisions, we did not consider the effects of droplet inertia, which, along with gravitational settling, can significantly influence the collision efficiency of larger droplet pairs. Additionally, it is well-established that cloud droplets usually possess surface charges, which necessitates the consideration of both external electric fields and the direct electrostatic interactions between charged droplets. Thus, a logical extension of this work involves incorporating exact hydrodynamic and electrostatic forces into the calculation of collision rates for inertial droplets settling in a laminar background flow. 

Here, we have assumed a deterministic background flow, but clearly atmospheric turbulence significantly influences the growth of cloud droplets and the initiation of precipitation \citep{shaw2003particle}. Specifically, for cloud droplets with radii in the range $15-40$ \textmu m, which fall within the so-called size gap regime, turbulence enhances collision rates of droplets by (i) increasing the relative radial velocities between droplet pairs \citep{saffman1956collision,falkovich2007sling}, and (ii) promoting the preferential concentration of inertial droplets in the straining regions of the turbulent flow \citep{sundaram1997collision,chun2005clustering}. Moreover, droplet hydrodynamic interactions further modulate collision processes \citep{pinsky2007collisions}. Recent direct numerical simulations coupled with Lagrangian particle tracking have provided compelling evidence that turbulence, along with droplet hydrodynamic interactions, play a pivotal role in broadening the droplet size distribution \citep{chen2018turbulence,michel2023influence}. However, there is a need to develop the theory of collision efficiencies for inertial droplets under the combined effects of turbulence, gravitational settling, and short-range and external electric field forces.

In this paper, we have concentrated solely on an idealized treatment of droplets that would be applicable to warm clouds, wherein the collision of water droplets underlies raindrop formation. In contrast, mixed-phase clouds contain both supercooled droplets and ice crystals, whose interactions significantly complicate precipitation processes. For example, whereas the aggregation of snow results from collisions between ice crystals, graupel grows by rimming, which occurs when descending ice crystals collide with supercooled droplets under turbulent conditions \citep{pruppacher1997microphysics,wang1994collision}. Unlike spherical droplets, due to their shape anisotropy and variable settling orientations, quantifying collisions involving ice crystals remains very challenging. Recent studies  have delved into the gravitational and turbulent dynamics affecting ice crystal collisions \citep{jucha2018settling,sheikh2022colliding}, as well as examining interactions between ice crystals and supercooled droplets \citep{naso2018collision,jost2019effect,sheikh2024effect}. However, it is noteworthy that these investigations typically employ the so-called ghost collision approximation, which neglects hydrodynamic and electrostatic interactions between colliding hydrometeors. Electrostatic forces become increasingly influential at small separation distances, considerably impacting the outcomes of ice crystal collisions. Recently, \citet{joshi2025electrostatic} quantitatively assessed the electrostatic forces and torques acting between charged anisotropic particles, revealing that these effects can lead to a preferential alignment of ice crystals, thereby affecting their collision dynamics. For a more thorough understanding of hydrometeor interactions, future research should seek to integrate these geometric and electrostatic influences into collision rate calculations.

\section*{ACKNOWLEDGMENTS}
P.P. and J.S.W.  gratefully acknowledge support from the Swedish Research Council under Grant No.638-2013-9243.

\appendix

\section{Formula for the collision efficiency for weak electric fields ($N_E \ll 1$) in the absence of van der Waals forces ($N_v = 0$).} \label{appendixA}

First, we derive a formula for the angle $\theta(r)$ used in the analysis of relative droplet trajectories under the influence of a weak electric field ($N_E \ll 1$). We assume $\theta(r) = \theta_0(r) + N_E \theta_1(r) + O(N_E^2)$, where $\theta_0$ and $\theta_1$ denote the leading-order behavior and first-order correction in $N_E$ respectively, and we ignore the van der Waals force ($N_v = 0$). With this ansatz, from equation (\ref{final_trajectory_equation}) the relative trajectory equations at O$(1)$ and O$(N_E)$ are
\begin{eqnarray}
    \frac{d\theta_0}{dr} &=& -\dfrac{3\left(1-B\right)\sin\theta_0 \cos\theta_0}{r\left(1-A\right) \left(3\cos^2\theta_0-1\right)}, \hspace{3mm} \text{and} \label{O(1)_trajectory_equation} \\ \frac{d\theta_1}{dr} &=& - T_1(r) \theta_1 + T_2(r), \label{O(NE)_trajectory_equation}
\end{eqnarray}
 where the terms $T_1(r)$ and $T_2(r)$ are given by
\begin{eqnarray}
   T_1(r) &=& \dfrac{3\left(1-B\right)\left(\cos 2\theta_0+3\right)}{2 r \left(1-A\right) \left(3\cos^2\theta_0-1\right)^2}, \label{T_1_term_order_1} \qquad\text{and} \\
   T_2(r) &=& \dfrac{3\left(1-B\right)G\kappa\sin 2\theta_0 \left(F_1\cos^2\left(\theta_0-\eta\right)+F_2\sin^2\left(\theta_0-\eta\right)\right)}{2 r^2 \left(1-A\right)^2 \left(3\cos^2\theta_0-1\right)^2} \nonumber \\ &+& \dfrac{\left(1-A\right)H\kappa F_8 \left(3\cos 2\theta_0+1\right)\sin\left(2\left(\eta-\theta_0\right)\right)}{2 r^2 \left(1-A\right)^2 \left(3\cos^2\theta_0-1\right)^2}. \label{T_2_term_order_NE} 
\end{eqnarray}
To calculate the upstream collisional area necessary for computing collision efficiency, we need to determine the limiting colliding trajectory. We can achieve this by imposing the condition $v_r = 0$, which allows us to determine the boundary conditions for equations (\ref{O(1)_trajectory_equation}) and (\ref{O(NE)_trajectory_equation}). Up to an arbitrary constant $C_1$, these boundary conditions for both the O$(1)$ and O$(N_E)$ trajectory equations are given by
\begin{eqnarray}
    &&\theta_0 (r = 2) = \arccos\left(\dfrac{1}{\sqrt{3}}\right), \hspace{3mm} \text{and} \label{O(1)_boundary_condition} \\
    &&\theta_1 = \left.\dfrac{G\kappa \left(F_1\cos^2\left(\theta_0-\eta\right)+F_2\sin^2\left(\theta_0-\eta\right)\right)}{6(1-A)r \sin\theta_0 \cos\theta_0}\right\vert_{r=2} = C_1. \label{O(NE)_boundary_condition}
\end{eqnarray}
Using these boundary conditions, the solutions for $\theta_0(r)$ and $\theta_1(r)$ are
\begin{eqnarray}
    \theta_0(r) &=& \cos^{-1} \left[\dfrac{2\times 3^{1/3} + 2^{1/3}\left(-9 I(r) + \sqrt{-12 + 81 I^2(r)}\right)^{2/3}}{6^{2/3}\left(-9 I(r) + \sqrt{-12 + 81 I^2(r)}\right)^{1/3}}\right], \hspace{3mm} \text{with} \hspace{3mm} I(r) = \dfrac{2}{3\sqrt{3}}\varphi(r), \label{Expression_for_theta0}\\
    \theta_1(r) & =& \exp \left(-\int_2^r T(r') dr'\right) \Bigg[C_1 + \int_2^r T_2(r') \exp \left(\int_2^{r'} T(r'') dr''\right)dr' \Bigg]. \label{Expression_for_theta1} 
\end{eqnarray}
The trajectory analysis in \S \ref{Results_and_discussion} suggests that upstream interception areas for weak electric fields are circles with their centers on the compressional axis. By evaluating $\theta_0(r)$ and $\theta_1(r)$ at a large separation distance, we can determine the non-dimensional radius of each upstream area, and the non-dimensional droplet relative velocity normal to these areas. These quantities allow us to evaluate the integral in Eq. (\ref{Collision_rate_upstream_interception_area}). Finally, we arrive at the following expression for the collision efficiency accurate to O$(N_E)$:
\begin{equation}
   E_{12} = \lim_{r \rightarrow \infty} \exp\left[\int_2^r\frac{3(B-A)}{r(1-A)}dr\right] \left[1+N_E\left(2\cot\theta_0(r)-\tan\theta_0(r)\right)\theta_1(r)\right] + O(N_E^2). 
\label{Analytical_expression_weak_field} 
\end{equation}

\eject 


\providecommand{\noopsort}[1]{}\providecommand{\singleletter}[1]{#1}%

\end{document}